# TEANet: A Transpose-Enhanced Autoencoder Network for Wearable Stress Monitoring on Resource-Constrained Device

Md Santo Ali, *Student Member, IEEE,* Sapnil Sarker Bipro, *Graduate Student Member, IEEE,* Mohammod Abdul Motin, *Senior Member, IEEE,* Sumaiya Kabir, *Member, IEEE,* Manish Sharma, *Senior Member, IEEE,* M. E. H. Chowdhury, *Senior Member, IEEE*

***Abstract*— Mental stress poses a significant public health concern due to its detrimental effects on physical and mental well-being, necessitating the development of continuous stress monitoring tools for wearable devices. Blood volume pulse (BVP) sensors, readily available in many smartwatches, offer a convenient and cost-effective solution for stress monitoring. This study presents a deep learning approach, a Transpose-Enhanced Autoencoder Network (TEANet), for stress detection using BVP signals on resource-constrained devices. The proposed TEANet model was trained and validated utilizing a self-developed RUET SPML dataset, and the publicly available wearable stress and affect detection (WESAD) dataset. It achieves the highest accuracy of 92.94% and 96.94%, F1 scores of 95.16% and 95.95%, and kappa of 0.8181 and 0.9350 for RUET SPML, and WESAD datasets, respectively. The proposed TEANet effectively detects mental stress through BVP signals with high accuracy, making it a promising tool for continuous stress monitoring. Furthermore, deploying the proposed model on the Raspberry Pi 3B+ enhances its potential for reliable real-time stress monitoring using resource-constrained devices.**

**Source Code: https://github.com/shantorueteee19/TEANet-ruet-spml**

## I. Introduction

STRESS is the organism's response to any change that induces physical or psychological strain and pressure, with the body's reactions, referred to as stress responses, being nonspecific and manifesting physiologically, behaviorally, and emotionally [1, 2]. Mental stress can manifest as physical and mental health issues, including cardiovascular disease, depression, and anxiety, making its recognition crucial due to the potential negative impact on overall health prompting individuals to monitor their stress levels in daily life. Stress detection involves analyzing facial expressions, text-based assessments, and physiological signals [1, 2] but the former methods are limited by variability, subjectivity, and external factors, making physiological signals more reliable for objective and real-time monitoring. A variety of physiological signals, including electrocardiogram (ECG), electromyogram (EMG), skin temperature (TEMP), BVP, electrodermal activity (EDA), acceleration (ACC), respiration (RESP), and electroencephalogram (EEG), have been extensively studied for stress monitoring, both individually and in combination, using diverse processing techniques and statistical or machine learning methods [1, 3].

While EEG effectively reflects brain activity changes during mental stress, its practical application is hindered by the inconvenience and cost of EEG headsets and electrodes placed on the body of subjects [4]. Similarly, EMG, requiring muscle attachment, is impractical for widespread use [5]. ECG provides a non-invasive way to monitor the electrical activity of the heart and detect stress-related changes in heart rate variability, but it is often expensive and cumbersome due to the need for multiple electrodes and specialized equipment [2, 6-8]. Commercial wearable devices primarily use BVP sensors for physiological monitoring due to their affordability [9]. Moreover, BVP has emerged as a viable alternative to ECG for physiological assessments, offering a non-invasive approach while effectively measuring heart rate and variability [10-15].

The use of machine learning (ML) and deep learning (DL) techniques for analyzing BVP signals has become a prominent approach in stress monitoring. Several studies [1, 16-27] have explored ML-based approaches for stress monitoring using BVP signals, either independently or in combination with other physiological signals. Traditional approaches typically rely on hand-crafted features, including temporal, frequency-domain, non-linear, and statistical features. However, traditional ML methods often rely on human expertise, known as feature engineering, which can be time-consuming, error-prone, and require domain knowledge. Additionally, these methods may struggle to capture complex, non-linear relationships within the data, especially as the dataset grows.

A plethora of DL-based models have been explored for mental stress assessment [3, 28-37]. Deep learning-based models, generally, outperform traditional machine learning-based models by effectively processing raw data and

Md Santo Ali, Sapnil Sarker Bipro, Mohammod Abdul Motin, and Sumaiya Kabir are with Department of Electrical & Electronic Engineering, Rajshahi University of Engineering & Technology, Bangladesh (Correspondence e-mail: m.a.motin@ieee.org).
Sumaiya Kabir is also with the School of Engineering, RMIT University, Melbourne, Australia (sumaiya.kabir.1993@ieee.org).
Manish Sharma is with the Electrical and Computer Science Engineering, Institute of Infrastructure, Technology, Research and Management (IITRAM), Ahmedabad, Gujarat, India (manishsharma@iitram.ac.in).
M. E. H. Chowdhury is with Electrical Engineering Department, Qatar University, Doha-2713, Qatar (mchowdhury@qu.edu.qa).

automatically extracting relevant features [38, 39]. Many studies utilize either hand-crafted features or images obtained by means of fast Fourier transform (FFT), short-term fast Fourier transform or some other joint time-frequency methods. The manual feature extraction process, however, may discard important information, leading to suboptimal performance. Additionally, converting time-series data into images can significantly increase computational complexity. When combined with deep learning, such approaches may limit the model's potential in real-time applications. Given the computational constraints and real-time requirements of wearable devices, we hypothesize that processing unimodal BVP data with deep learning models installed in wearable devices can be the optimal approach for stress assessment.

This study introduces a transpose-enhanced autoencoder network for stress monitoring from BVP signals. The primary objective of this research is to develop a high-performance unimodal DL-based solution for real-world stress monitoring in resource constraint device. The key contributions of this work are as follows:

- A novel transpose-enhanced autoencoder network (TEANet) model for stress monitoring using wearable device-derived BVP signals, capable of delivering balanced performance on highly imbalanced datasets while outperforming existing state-of-the-art approaches.
- A sliding window segmentation-based augmentation strategy is implemented to address the class imbalance problem.
- A private dataset was developed for this study, and extensive model validation was conducted using both the publicly available dataset and the newly created private dataset.
- Furthermore, the deployment of the proposed model in a resource-constrained environment, demonstrating its practical feasibility for real-time stress monitoring in wearable applications.

This paper is organized as follows: Section II describes the materials and methods, Section III presents the results and discussion, and Section IV concludes the proposed study.

## II. Materials And Methods

### A. Dataset Description

In this study, we validated the proposed model using two datasets to distinguish between normal and stressed states. First, we collected physiological data from 22 participants, referred to as the RUET SPML Stress Dataset. Due to sensor malfunction, data from three participants were excluded, resulting in 19 participants (15 males and 4 females; mean age: 24 ± 2.8 years) for analysis. The experimental protocol consisted of a 3-minute baseline period followed by a Stroop test, simple problem-solving tasks, and a Sudoku puzzle. Signals were recorded using the Empatica E4 wristband. In addition, we utilized the publicly available WESAD dataset [40], which includes data from 15 participants (12 males and 3 females; mean age: 27.5 ± 2.4 years) under baseline, stress, and amusement conditions. Validating the model on both private and public datasets strengthens its robustness and generalizability.

### B. Signal Pre-Processing

The data were segmented into 30-second non-overlapping windows and paired with their corresponding labels. To ensure the signal amplitudes remained within a specific range, BVP signals were z-score normalized [2, 39].

The RUET SPML dataset has more stress recordings, while the WESAD dataset has a larger baseline class, leading to class imbalances that affect both datasets. To address this, data augmentation was applied during training. We adopted a simple sliding window-based method to augment the minority class. Initially, a non-overlapping 30-second windows were used to extract segments from the minority class. The data points and samples for both classes were calculated to balance the minority class. Minority windows were merged into a continuous signal, from which augmented segments were generated. The step size d, defined as the difference between successive indices, was determined using equation (1). The augmentation process is explained in Algorithm 1.

$$d = \frac{(t_{minority} - t_{window}) \times F_s}{n_{majority} - 1} \quad (1)$$

*where,*

$t_{minority}$ = *time length of minority class*
$t_{window}$ = *time length default window*
$F_s$ = *sampling frequency of the signal*
$n_{majority}$ = *number of samples of majority class*

**Algorithm 1:** Generation of augmented minority class segments using a 30-second sliding overlapping window technique, ensuring an equal number of samples between the minority and majority classes to address data imbalance effectively.

**Data:**

- *A data frame containing the minority class with desired window length.*
- *Parameters: desired window (*$t_{window}$ *in* sec*), sampling frequency of the signal (*$F_s$ *in* Hz*), number of augmented samples to generate for balancing*

**Result:**

*Augmented minority class training set with* $n_{majority}$ *samples.*

**Data Augment:**

**Start**

1 *Reshape the minority class windowed data frame to signal, i.e., the whole signal into a single row, and compute the step size (d) between two windows from equation (1).*
2 *Initialize start_index = 0 and end_index =* $t_{window} \times F_s$*.*
3 *Generate augmented samples:*
   *Repeat* $n_{majority}$ *times:*
      *a) Extract segment from reshaped data and append to augmented dataframe.*
      *b) Update indices:*
   *start_index = start_index + d*
   *end_index = end_index + d*
4 *Append minority class labels to all rows of the augmented data frame.*
5 *Return augmented data frame.*

**End**

### C. Transpose-Enhanced Autoencoder Network (TEANet)

For stress detection, the proposed transpose-enhanced autoencoder network architecture, demonstrated in Fig. 1, consists of an input layer connected to a down-sampling block, followed by transpose-enhanced autoencoder layers and classification blocks with a view to extracting various features from the BVP signals. The transpose-enhanced autoencoder (TEA) layer comprises two parallel paths: the transpose-enhanced (TE) path and the convolutional path. The transpose-enhanced path incorporates a transposed convolutional block, convolutional blocks, and autoencoder blocks, whereas the convolutional path comprises two convolutional blocks. The outputs of these paths are concatenated and forwarded to the classification blocks, thus passed to the classification stage. Let, the feature map from the dataset be $X \in \mathbb{R}^{N \times d}$ and each sample $x_i \in X$, where, $x_i = \{x_1, x_2, x_3, \ldots, x_N\}$, and N represents the total number of samples, and d represents data points. However, detailed descriptions of the structure and functionality of each block are provided below.

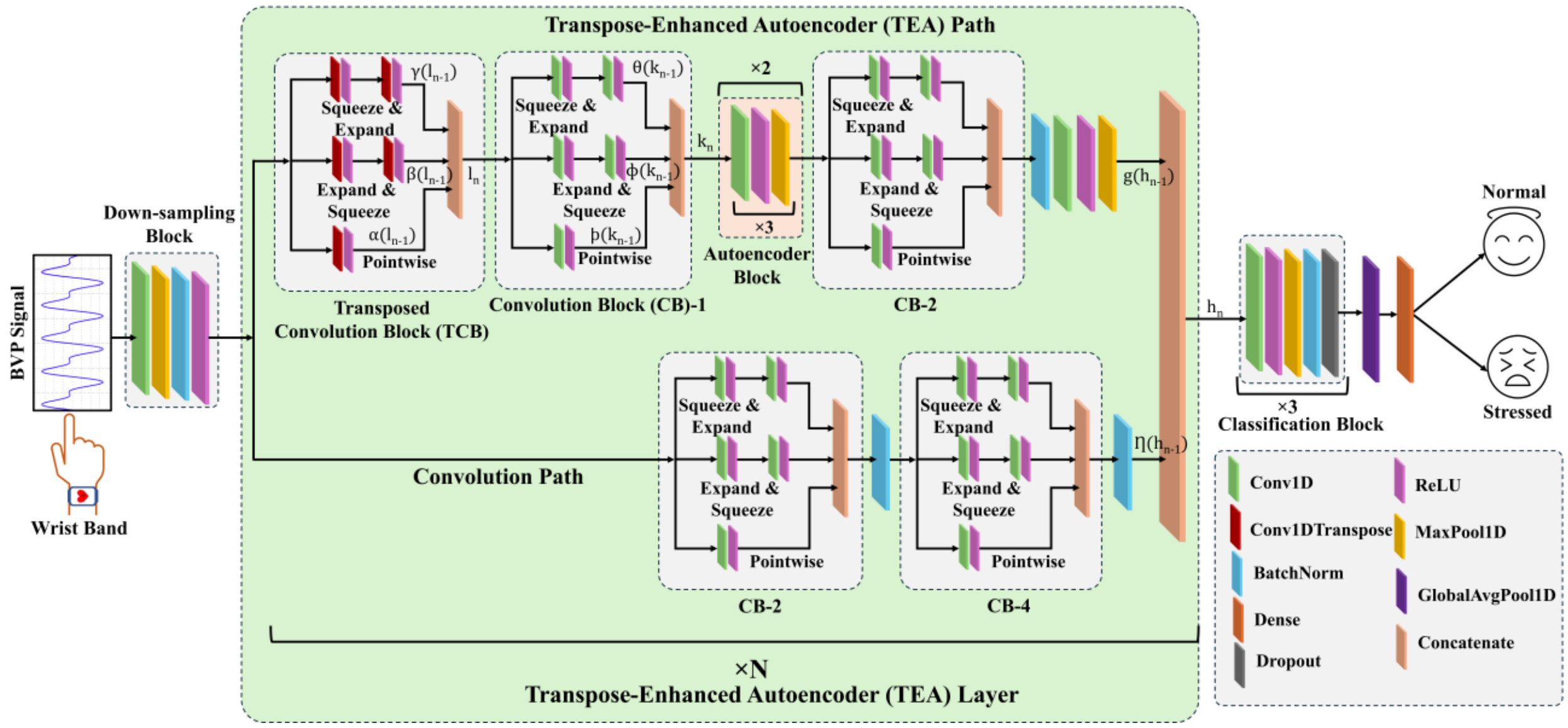

Fig. 1. The detailed architectural framework of the transpose-enhanced autoencoder network (TEANet): a novel approach for addressing class imbalance and enhancing robustness in BVP-based stress detection for wearable devices.

### 1) *Down-sampling Block (DSB)*

This block consists of a Conv1D layer with 128 filters, a kernel size of 5, and a stride of 4, followed by MaxPooling1D with a pool size of 2, stride of 2, batch normalization, and ReLU activation. This block reduces the input length to one-eighth in two steps: one-fourth via Conv1D and half via MaxPooling1D, which reduces the complexity in the subsequent layers. It also extracts key features in the initial stages, making it an essential component for efficient feature representation. Mathematically the outputs from the Conv1D layer,

$$z_i[j] = \sum_{t=0}^{k-1} w[t] * x_i[j+t] + b \quad (2)$$

where, $j = 1,2,3, \ldots, \frac{d-k}{s} + 1$, k = filter size, $s$ = strides, $w$ = weights, $b$ = biases.

The output from the MaxPooling1D layer,

$$z_j^{pool}[j] = \max\{z_i[j.s_p], z_i[j.(s_p+1)], \ldots, z_i[j.s_p + p - 1]\} \quad (3)$$

where, $s_p$ = strides, $p$ = pool size.

The output from the BatchNormalization layer,

$$\hat{z}_i[j] = \frac{z_j^{pool}[j] - \mu_j}{\sqrt{\sigma_j^2 + \varepsilon}} \quad (4)$$

where, $\mu_j = mean$, $\sigma_j = standard\ deviation$, $\varepsilon$=*a small positive number to prevent division by zero*. After applying learnable parameters $\delta$ (scale) and $\Delta$ (shift),

$$\tilde{z}_i[j] = \delta.\hat{z}_i[j] + \Delta \quad (5)$$

The ReLU introduces nonlinearity followed by the equation (6).

$$o_i[j] = \max(0, \tilde{z}_i[j]) \quad (6)$$

The output, $o_i[j]$ then fed to the TEA layer.

### 2) *Transpose-Enhanced Autoencoder (TEA) Layer*

The TEA layer consists of fundamental building blocks namely convolutional block (CB), transpose convolutional block (TCB), and autoencoder block. The fundamental blocks are detailed below.

#### *a) Convolutional Block (CB)*

The convolutional block comprises three distinct paths, each contributing uniquely to feature-extraction and transformation. The pointwise path *þ(.)*, consisting of a single Conv1D layer with a kernel size of 1 followed by ReLU activation, plays a crucial role in transforming input features into a new feature space by focusing on inter-channel dependencies. In contrast, the expand-and-squeeze path $\phi(.)$ and squeeze-and-expand path $\theta(.)$ utilize Conv1D layers with kernel sizes of 9 and 3, respectively, each followed by ReLU activation. The expansion step enhances the receptive field, capturing broader contextual information, while the squeezing step refines the feature representation by focusing on local patterns. At the end of this block, the outputs from all three paths are concatenated, effectively combining diverse feature representations to enrich the input for subsequent layers. The features $k_{n-1}$ from the previous block are processed to provide the output $k_n$ as described in equation (7).

$$k_n = concatenate[\theta(k_{n-1}), \phi(k_{n-1}), \text{þ}(k_{n-1})] \quad (7)$$

#### *b) Transpose Convolutional Block (TCB)*

This block is designed similarly to the convolutional clock, with the primary distinction being the replacement of the Conv1D layer by the Conv1DTranspose layer. The pointwise transpose convolution operation $\alpha(.)$ applies learnable weights to perform a feature-space transformation without altering the temporal resolution of the input. In contrast, the expand-and-squeeze path $\beta(.)$ captures broader temporal patterns, while the squeeze-and-expand $\gamma(.)$ path focuses on local patterns of the input features. The hierarchical features extracted from these three paths are aggregated using a concatenate layer. This block enhances high-level feature representations by enabling the model to reconstruct information from compressed feature maps. It ensures the recovery of fine-grained details that may have been lost during the down-sampling process in earlier TEA layers, especially when multiple TEA layers are employed. Furthermore, the multi-scale reconstructions provided by this block enhance robustness in handling complex input patterns, such as those encountered in stress monitoring tasks. The input features from the previous block, $l_{n-1}$, are processed to compute the output $l_n$, as outlined in Equation (8).

TABLE I
THE NUMBER OF FILTERS IN THE TRANSPOSE-ENHANCED AUTOENCODER (TEA) LAYERS

| Block | Transposed Conv Block | | | Conv Block-1 | | | Conv Block-2 | | | Conv Block-3 | | | Conv Block-4 | | |
|---|---|---|---|---|---|---|---|---|---|---|---|---|---|---|---|
| TEA-N Layers | $f_p$ | $f_{ese}$ | $f_{ses}$ | $f_p$ | $f_{ese}$ | $f_{ses}$ | $f_p$ | $f_{ese}$ | $f_{ses}$ | $f_p$ | $f_{ese}$ | $f_{ses}$ | $f_p$ | $f_{ese}$ | $f_{ses}$ |
| 1 | 64 | 64 | 16 | 16 | 32 | 64 | 16 | 32 | 32 | 16 | 32 | 96 | 16 | 96 | 96 |
| 2 | 32 | 32 | 16 | 16 | 32 | 64 | 16 | 96 | 96 | 16 | 96 | 96 | 16 | 64 | 64 |
| 3 | 64 | 64 | 32 | 16 | 32 | 64 | 16 | 96 | 96 | 16 | 16 | 32 | 16 | 64 | 64 |
| 4 | 64 | 64 | 16 | 16 | 32 | 96 | 16 | 16 | 32 | 16 | 32 | 64 | 16 | 96 | 96 |
| 5 | 64 | 64 | 96 | 64 | 16 | 16 | 16 | 32 | 32 | 16 | 64 | 64 | 64 | 96 | 96 |
| 6 | 64 | 64 | 96 | 64 | 16 | 16 | 16 | 32 | 32 | 16 | 64 | 64 | 64 | 96 | 96 |
| 7 | 64 | 64 | 96 | 64 | 16 | 16 | 16 | 32 | 32 | 16 | 64 | 64 | 64 | 96 | 96 |

$$l_n = concatenate[\alpha(l_{n-1}), \beta(l_{n-1}), \gamma(l_{n-1})] \quad (8)$$

***c) Autoencoder Block (AB)***

The autoencoder block is used for feature extraction through convolution operations. It consists of three Conv1D layers, each followed by MaxPooling1D with a pool size of 2. The convolutional layers have kernel sizes of 3 and filters of 96, 16, and 32, respectively. This block down-samples the features while emphasizing important ones, enhancing the model's ability to capture key patterns.

The blocks described above are the building blocks for the TEA layer. As earlier described, this layer comprises two main components including the convolutional path $\eta(.)$, and the transpose-enhanced (TE) path $g(.)$. These components work together to extract and transform features, with the Convolutional path capturing local and global patterns, while the TE path reconstructs important features. The outputs from both paths are then concatenated and passed to the classification blocks for final processing. The descriptions of two paths are detailed below

***i. Convolutional Path, η(.)***

The convolutional path consists of two convolutional blocks each followed by batch normalization, designed to extract and refine features through successive convolutional operations. These blocks leverage multiple Conv1D layers with varying kernel sizes and ReLU activations to capture both local and global patterns in the input data. By combining these blocks, the Convolutional Path effectively enhances the feature representation, providing richer inputs for the subsequent layers of the model.

***ii. Transpose-Enhanced (TE) Path, g(.)***

The transpose-enhanced path consists of a transpose-enhanced block, several convolutional blocks, and two autoencoder blocks. The transposed convolutional block reconstructs key features from previous layers, which are then passed through a convolutional path. This path propagates the features through an autoencoder block, followed by a convolutional block, batch normalization, another Conv1D layer for shape matching with kernel size 1, and a ReLU activation. Finally, the features are downsampled with a MaxPooling1D layer (pool size 2). This path effectively reconstructs and refines important features for subsequent processing. If the previous features are represented as $h_{n-1}$, then the output,

$$h_n = concatenate[\eta(h_{n-1}). g(h_{n-1})] \quad (9)$$

The output $h_n$ from the TEA layer is then fed to the classification block for further processing.

***3) Classification Block***

The classification block consists of Conv1D, ReLU, MaxPooling1D, Batch Normalization, and Dropout layers, with a pool size of 2 and a dropout rate of 0.3. This block is repeated three times, with the number of filters changing from 96 to 64, and then to 32. Afterward, a global average pooling layer processes the concatenated output, and the final prediction is generated by a dense layer with a softmax activation function, described as: $softmax(\boldsymbol{x}) = \frac{e^{x_i}}{\sum_{j=1}^{n} e^{x_j}}$ (10)

To facilitate a comparison of the performance of the proposed architecture, we evaluate configurations with two to seven layers. The filter numbers used in the TEA Layers, which have not been previously specified, are detailed in Table I for clarity and reference. The numbers of filters are denoted *by* $f_p$, $f_{ese}$, $f_{ess}$, $f_{ses}$, $f_{see}$, representing the number of filters in the pointwise layer, expand layer of the expand-and-squeeze path, squeeze layer of the expand-and-squeeze path, squeeze layer of the squeeze-and-expand path, and expand layer of the squeeze-and-expand path, respectively. We have chosen $f_p = f_{ess} = f_{see}$.

Root Mean Square Propagation (RMSprop) is an optimization algorithm designed to handle the challenges of training deep neural networks [41], particularly dealing with the problem of vanishing, and exploding gradients by adapting the learning rate based on recent gradients, was used as the optimizer. It computes the exponentially weighted moving average of the squared gradients as the equation (11),

$$E[g^2]_t = \gamma E[g^2]_{t-1} + (1-\gamma)g_t^2 \quad (11)$$

where $\gamma$ is the decay rate, $g$ represents the gradient.

And updates the parameters as defined by the equation (12).

$$\theta_{t+1} = \theta_t - \frac{\rho}{\sqrt{E[g^2]_t + \varepsilon}} g_t \quad (12)$$

where $\theta$ is the learnable parameters, $\rho$ is the learning rate, and $\varepsilon$ is a small constant to prevent division by zero.

In this study, a sparse categorical cross-entropy loss function was used, which is often applied in multi-class classification tasks but can also be used for binary classification, measures the difference between two probability distributions: predicted probabilities $\hat{y}$ and actual labels $y$. This loss function $L(y, \hat{y})$, described by Equation (13), was employed to calculate the loss during training.

TABLE II
PERFORMANCE COMPARISON OF THE TRANSPOSE-ENHANCED AUTOENCODER NETWORK WITH VARYING NUMBERS OF TEA LAYERS. THE BEST RESULTS ARE HIGHLIGHTED IN BOLD. THE NUMBER OF PARAMETER COUNTS (#P) OF EACH MODEL IS REPORTED IN MILLIONS (M).

| **Dataset** | **RUET SPML** | | | | | | **WESAD** | | | | | | **#P** |
|---|---|---|---|---|---|---|---|---|---|---|---|---|---|
| **TEA-N Layers** | **Acc** | **Spe** | **Sen** | **F1** | **AUC** | **k** | **Acc** | **Spe** | **Sen** | **F1** | **AUC** | **k** | |
| **TEA-7 Layers** | 91.20 | 83.33 | 93.84 | 94.04 | 0.8925 | 0.7672 | 95.84 | 96.03 | 95.40 | 94.34 | 0.9814 | 0.9105 | 4.17 |
| **TEA-6 Layers** | 91.19 | 84.21 | 93.55 | 94.02 | 0.9131 | 0.7667 | 96.29 | 96.55 | 95.71 | 95.03 | 0.9807 | 0.9208 | 3.13 |
| **TEA-5 Layers** | 92.51 | 82.46 | 95.89 | 95.03 | 0.9261 | 0.7915 | 95.85 | 96.37 | 94.79 | 94.46 | 0.9803 | 0.9114 | 2.08 |
| **TEA-4 Layers** | 91.18 | 81.58 | 94.43 | 94.08 | 0.9166 | 0.7621 | 96.58 | 97.06 | 95.71 | 94.92 | **0.9849** | 0.9239 | 1.43 |
| **TEA-3 Layers** | **92.94** | **88.60** | 94.43 | **95.16** | **0.9384** | **0.8181** | **96.94** | **97.06** | **96.63** | **95.95** | 0.9824 | **0.9350** | 1.11 |
| **TEA-2 Layers** | 92.07 | 83.33 | **95.01** | 94.79 | 0.9190 | 0.7697 | 95.09 | 94.30 | 96.32 | 93.89 | 0.9751 | 0.8984 | 0.78 |

$$L(y, \hat{y}) = -\sum_{c=0}^{1} y_c \log(\hat{y}_c) \tag{13}$$

It is particularly suitable for binary classification tasks, helping the model to differentiate between the two classes more effectively.

## D. *Performance Metrices*

We have evaluated several indices for stress classification (normal versus stressed) performance measurement of the TEANet including accuracy, sensitivity, specificity, F1-score, area under the curve (AUC) and Cohen's kappa coefficient (k). To rigorously evaluate the proposed model, this study employed leave-one-subject-out (LOSO) cross-validation. In this method, one subject is reserved for testing, while the model is trained in the remaining subjects. This iterative process ensures that each subject is used as a test case exactly once, providing a comprehensive assessment of the model's generalizability and performance across diverse individuals.

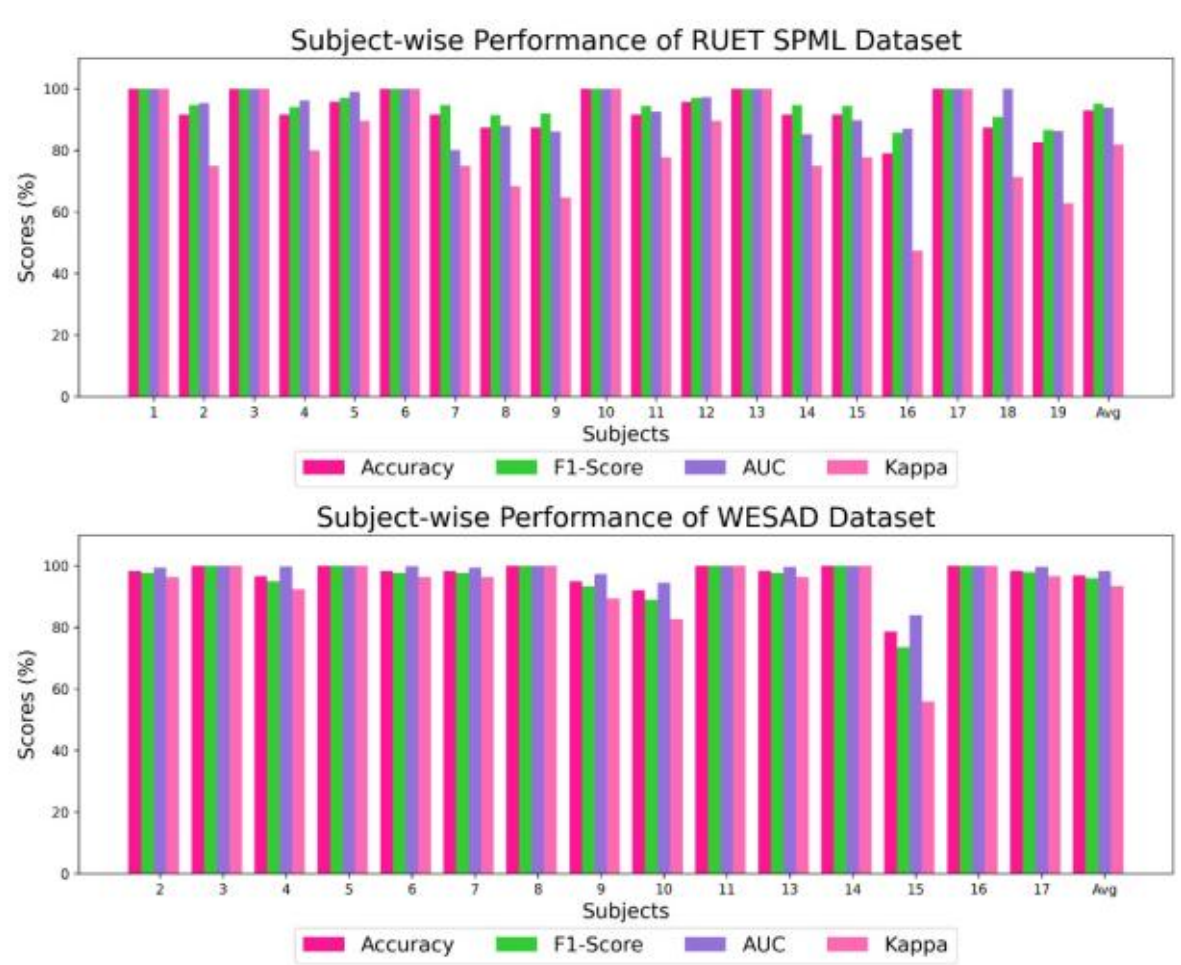

Fig. 2. Subject-wise performances including accuracy in %, F1-score in %, AUC, and Cohen's k values of, RUET SPML dataset (top) and WESAD dataset (bottom), both with TEA-3 layers.

# III. RESULTS AND DISCUSSION

The performance of TEANet was evaluated across varying TEA layers, denoted as TEA-N layers, ranging from TEA-2 to TEA-7, where N represents the number of TEA layers. As shown in Table II, the model achieves the highest overall performance on both the RUET SPML and WESAD datasets when configured with TEA-3 layers. Based on this observation, TEA-3 was selected for detailed analysis on both datasets. Furthermore, the parameter counts increase progressively with the number of TEA layers (i.e., as N increases in TEA-N). Despite this, TEANet with TEA-3 achieves the best performance while maintaining 1.11M parameters, indicating an optimal balance between accuracy and model complexity. The proposed model demonstrates remarkable performance, achieving 92.94% accuracy, 88.60% specificity, 94.43% sensitivity, 95.16% F1-score, 0.9384 AUC, and 0.8181 kappa with the RUET SPML dataset. On the other hand, 96.93% accuracy, 97.06% specificity, 96.63% sensitivity, 95.95% F1-score, 0.9824 AUC, and 0.9350 kappa score with the WESAD dataset. Fig. 2 highlights the variations in performance across subjects for both datasets. The results underscore the model's outstanding capability to distinguish between normal and stressed classes. Moreover, the performance across TEA-2 to TEA-7 layers shows minimal deviation, highlighting the scalability and optimization potential of TEANet, making it a promising solution for stress detection in diverse real-world scenarios.

## A. *Effect of Augmentation*

Data augmentation was applied during training to reduce the model bias toward the majority class, effectively decreasing the likelihood of overfitting [42]. The effect of data augmentation on the model performance is demonstrated in Table III. For the RUET SPML dataset, data augmentation improves accuracy by 5.70%, specificity by 29.83%, F1-score by 2.99%, AUC by 14.05%, and kappa by 25.63%, while slightly reducing sensitivity by 2.34%. For the WESAD dataset, data augmentation enhances accuracy by 3.05%, specificity by 1.55%, sensitivity by 5.83%, F1-score by 4.55%, AUC by 1.55%, and kappa by 6.68%. Although sensitivity decreases by 1.17% in the RUET SPML dataset, the overall performance, particularly for the minority class, improves significantly due to data augmentation. These results underscore the effectiveness of data augmentation in improving minority class accuracy, i.e., specificity in RUET SPML dataset, and sensitivity in WESAD dataset, leading to a more balanced overall performance.

TABLE III
PERFORMANCES OF THE TEANET WITH AND WITHOUT AUGMENTATION TO STUDY THE EFFECT OF AUGMENTATION.

| **Dataset** | **Data Aug** | **Acc** | **Spe** | **Sen** | **F1** | **AUC** | **k** |
|---|---|---|---|---|---|---|---|
| **RUET SPML** | **No** | 87.24 | 58.77 | **96.77** | 92.17 | 0.7979 | 0.5618 |
| | **Yes** | **92.94** | **88.60** | 94.43 | **95.16** | **0.9384** | **0.8181** |
| **WESAD** | **No** | 93.89 | 95.51 | 90.80 | 91.40 | 0.9669 | 0.8682 |
| | **Yes** | **96.94** | **97.06** | **96.63** | **95.59** | **0.9824** | **0.9350** |

## B. *The Effect of Segment Length on Model Performance*

The influence of segment length on model performance was systematically evaluated across both datasets. Fig. 3 presents the results of the proposed model using segment lengths varying from 10 to 40 seconds. Notably, the 30-second segments yielded the highest performance in terms of accuracy, Cohen's kappa, and minority-class classification for both datasets. These results indicate that a 30-second segment length offers an optimal trade-off between contextual richness and computational efficiency, making it suitable for practical stress detection applications. Conversely, the model exhibited its lowest performance with 10-second segments, likely due to the limited contextual information available in shorter durations, which adversely affects model's performance. Although longer segments, such as 40 seconds, capture more stress-related physiological information, they increase computational complexity and reduce the number of training samples, which in turn leads to diminished overall performance.

TABLE IV
THE COMPARISON OF PERFORMANCES WITH THE DIFFERENT VARIANTS OF THE PROPOSED ARCHITECTURE AND THE SELECTION PROCESS OF THE TEANET. THE BEST RESULTS ARE HIGHLIGHTED IN BOLD.

| Dataset | RUET SPML | | | | | | WESAD | | | | | |
|---|---|---|---|---|---|---|---|---|---|---|---|---|
| **Model** | **Acc** | **Spe** | **Sen** | **F1** | **AUC** | **k** | **Acc** | **Spe** | **Sen** | **F1** | **AUC** | **k** |
| **TE Path without TCB** | 86.57 | 59.65 | 95.60 | 91.62 | 0.8441 | 0.5650 | 94.41 | 95.85 | 91.72 | 91.49 | 0.9644 | 0.8770 |
| **TE Path without Autoen.** | 88.76 | 63.16 | **97.36** | 93.40 | 0.7984 | 0.6339 | 95.94 | 98.45 | 91.41 | 93.40 | 0.9788 | 0.9062 |
| **TE Path** | 87.66 | 67.54 | 94.43 | 92.06 | 0.8390 | 0.6290 | 94.75 | 94.30 | 95.40 | 93.18 | 0.9721 | 0.8895 |
| **Conv Path** | 91.39 | 75.44 | 96.77 | **94.51** | 0.9105 | 0.7387 | 95.31 | 95.85 | 94.17 | 93.39 | 0.9738 | 0.8981 |
| **TEANet** | **92.94** | **88.60** | 94.43 | **95.16** | **0.9384** | **0.8181** | **96.94** | **96.06** | **95.63** | **95.95** | **0.9824** | **0.9350** |

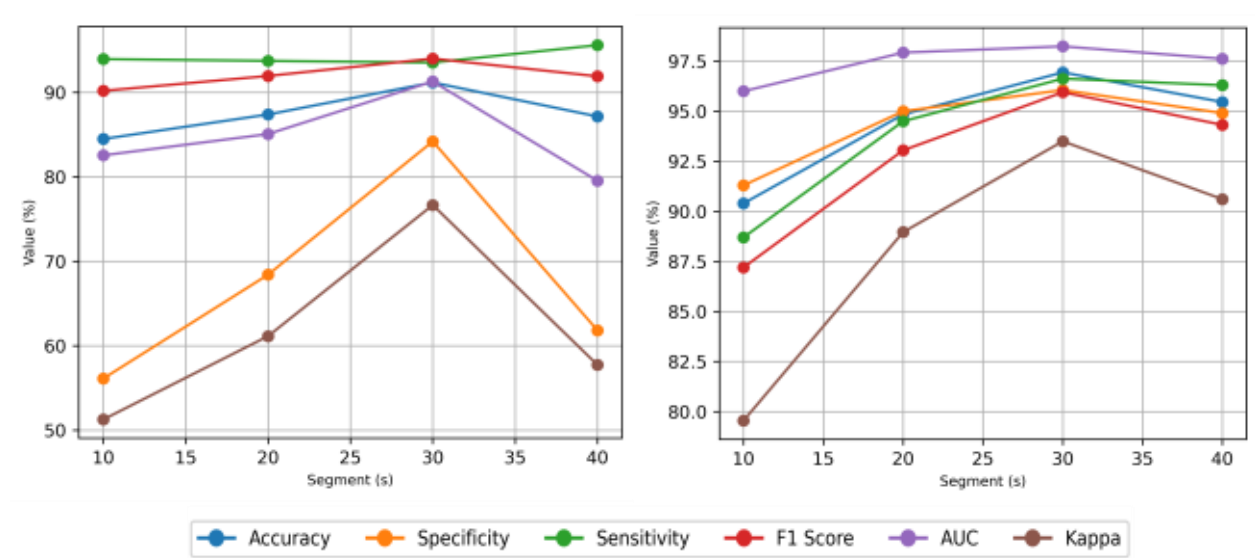

Fig. 3. Segment-wise trade-off analysis for the RUET SPML dataset (left) and the WESAD dataset (right). The analysis indicates that a segment length of 30 seconds offers an optimal balance between performance and computational efficiency.

## C. Ablation Study

Our proposed TEA layer comprises two components: the TE path and the convolutional path. The TE path includes a TCB, CB, autoencoder block, and supporting layers. To evaluate their contributions, an ablation study was done considering four TEANet variants:

1. TE path without TCB
2. TE path without autoencoder
3. TE path only
4. Convolutional path only

The TE path without the TCB module showed biased performance toward majority class in both datasets, predicting the minority class (i.e., specificity) of 59.6% for RUET SPML and 91.7% (sensitivity) for WESAD. Excluding the autoencoder module, i.e., utilizing the TE path without the autoencoder, improves overall accuracy, particularly with the RUET SPML dataset. However, it offers limited performance on the minority-class for both datasets, which remains our primary focus. On the RUET SPML dataset, specificity was 67.5%, while on WESAD, sensitivity was 95.4%. Comparatively, the full TE path achieved minority class accuracies of 63.2% (RUET SPML) and 91.4% (WESAD). These results demonstrate the autoencoder's role in mitigating class imbalance. The convolutional path alone outperformed individual TE path variants, achieving 75.4% and 94.2% accuracies for the minority class in RUET SPML and WESAD datasets, respectively. However, neither path independently achieved balanced performance. The combined paths, the full TEANet architecture, delivered more consistent results for both the datasets. The results of the ablation study are shown in Table IV. In summary, the inclusion of the autoencoder module and the integration of TEA and convolutional paths, i.e., the complete TEANet were essential for addressing class imbalance. This study underscores the effectiveness of the proposed architecture in handling imbalanced datasets, particularly the challenging RUET SPML dataset.

## D. Comparison with State-of-the-Art Architecture

To assess the efficiency of our proposed TEANet model, a comparative analysis with established convolutional and transformer-based architectures was performed. The performance of our model was compared with AlexNet, MobileNet-V1, and ResNet-18 from convolutional models, as well as CNN with transformer (CNN+TF) and multi-perspective channel-attention with transformer (MPCA+TF) [43] from transformer-based models. While some of these models demonstrated moderate performance on the WESAD dataset, they struggled significantly to address class imbalance in the RUET SPML dataset and failed to maintain balanced performance across both datasets. In contrast, the proposed TEANet consistently outperformed all the state-of-the-art architectures. Notably, TEANet achieves this with only 1.11M parameters and a model size of 9.24 MB making it 2.75 to 72 times smaller in parameter count and 2.54 to 66 times smaller in size compared to competing models, as shown in Table V. These results highlight TEANet's superior efficiency and suitability for deployment in resource-constrained wearable devices.

TABLE V
PERFORMANCE COMPARISON OF THE TRANSPOSE-ENHANCED AUTOENCODER NETWORK WITH STATE-OF-THE-ART ARCHITECTURE. THE BEST RESULTS ARE HIGHLIGHTED IN BOLD.

| Type | Model | Dataset | | | | | | | | | | Parameters | |
|---|---|---|---|---|---|---|---|---|---|---|---|---|---|
| | | RUET SPML | | | | | WESAD | | | | | | |
| | | **Acc** | **Spe** | **Sen** | **AUC** | **k** | **Acc** | **Spe** | **Sen** | **AUC** | **k** | **#P** | **Size** |
| **Convolutional** | **AlexNet** | 81.96 | 42.11 | 95.31 | 0.6425 | 0.3874 | 94.53 | **96.37** | 91.10 | 0.9580 | 0.8778 | 79.82 | 609 |
| | **MobileNet-V1** | 80.43 | 29.82 | 97.36 | 0.6465 | 0.2919 | 95.59 | 95.51 | 95.71 | 0.9748 | 0.9032 | 3.05 | 23.5 |
| | **ResNet-18** | 90.07 | 71.05 | 96.48 | 0.8673 | 0.6901 | 96.71 | 95.68 | **98.47** | 0.9872 | 0.9315 | 5.26 | 40.3 |
| **Transformer** | **CNN+TF** | 85.03 | 51.75 | 96.19 | 0.7611 | 0.4822 | 92.52 | 95.51 | 87.12 | 0.9329 | 0.8314 | 4.86 | 37.2 |
| | **MPCA+TF** | 78.23 | 13.16 | **100.00** | 0.5103 | 0.1368 | 95.31 | 95.68 | 94.48 | 0.9634 | 0.9001 | 6.23 | 47.7 |
| **This Work** | **TEANet** | **92.94** | **88.60** | 94.43 | **0.9384** | **0.8181** | **96.94** | 96.06 | 96.63 | **0.9824** | **0.9350** | **1.11** | 9.24 |

### *E. Comparative Analysis with the State-of-the-Art-Models for Stress Monitoring Using Wearable Device*

A comparative analysis of our work with existing BVP-based stress classification studies is summarized in Table VI. This comparative analysis serves to contextualize our contributions within the broader field and highlight the novel aspects of our proposed approach. The comparative analysis presented in Table VI reveals that our proposed method outperforms the majority of existing BVP-based stress classification studies, with notable exceptions being studies [1], [19], and [21]. A common characteristic these works, including [1], [19], and [21], is their reliance on feature engineering techniques. Studies [1], [21], and [44] further complicate its methodology by incorporating multiple physiological signals and relying on manual feature extraction in traditional machine learning approaches. While the study [19] utilizes only BVP signals, it introduces a complex peak detection and denoising process followed by traditional ML classifiers. In contrast, our method achieves competitive performance by directly processing raw BVP signals without the need for intricate preprocessing or handcrafted features. This streamlined approach enhances the practical applicability and generalizability of our model.

TABLE VI
THE COMPARATIVE ANALYSIS OF OUR WORK WITH THE EXISTING LITERATURES FOCUSING ON STRESS MONITORING

| Ref. | Dataset | Signals | Method | Acc. |
|---|---|---|---|---|
| Alshamrani [36] | WESAD | BVP, ACC, TEMP | FCN | 85% |
| Bellante *et al.* [27] | WESAD | BVP, EDA, RESP | SVM | 97.2% |
| Bobade *et al.* [1] | WESAD | ACC, ECG, BVP, EDA, EMG, RESP | ANN | 95.21% |
| Benchekroun *et al.* [23] | Private | BVP | RF | 83% |
| Rashid *et al.* [4] | WESAD | BVP | H-CNN | 88.56% |
| Zhu *et al.* [28] | WESAD | BVP | RF | 78.8% |
| Schmidt *et al.* [3] | WESAD | BVP | LDA | 85.83% |
| Heo *et al.* [25] | WESAD | BVP | LDA | 96.5% |
| Jahanjoo *et al.* [54] | WESAD | BVP | SVM | 95.55% |
| Kalra *et al.* [38] | Private | BVP | DNN | 90.5% |
| **This work** | RUET SPML | **BVP** | **TEANet** | **92.9%** |
| **This work** | WESAD | **BVP** | **TEANet** | **96.9%** |

The proposed TEANet model excels in stress classification. The study [1] reports 95.21% accuracy, but with a large number of multimodal signals; on the other hand, Heo *et al.* [19] achieved 96.5% accuracy with LDA, but with a larger window length of 120 seconds. Additionally, Jahanjoo *et al.* [44] achieve a good accuracy of 95.55% with a larger window size, and Bellante *et al.* [21] report better performance than ours, but also with larger windows and a larger number of signals, unlike ours, which relies solely on BVP. A thorough comparison with existing methods emphasizes the innovative approach and enhanced performance of TEANet, establishing it as a highly effective solution for real-world stress monitoring, particularly in wearable technology applications.

### *F. Poor-quality Signal and Effect on Performance*

The proposed model demonstrates robust performance across the majority of subjects; however, reduced accuracy is observed for certain subjects, primarily due to limitations in signal quality. For instance, Subject 19 from the RUET SPML dataset achieves an accuracy of 79%, while Subject 15 from the WESAD dataset reaches 83%. Further analysis of these cases highlights inherent challenges associated with BVP signal acquisition. As illustrated in Fig. 4, the BVP signal of Subject-19 is heavily contaminated by motion artifacts, while the signal from Subject-15 exhibits both pronounced motion-induced distortions and attenuated cardiac waveform features. These physiological and technical constraints represent inherent limitations of BVP-based monitoring that impact all analytical methods. The model's performance on these subjects reflects the maximum achievable accuracy given the available signal quality rather than architectural limitations. These results underscore the critical need for improved signal acquisition and preprocessing techniques in wearable applications, where robustness in handling motion artifacts remains a significant and ongoing research challenge in the field.

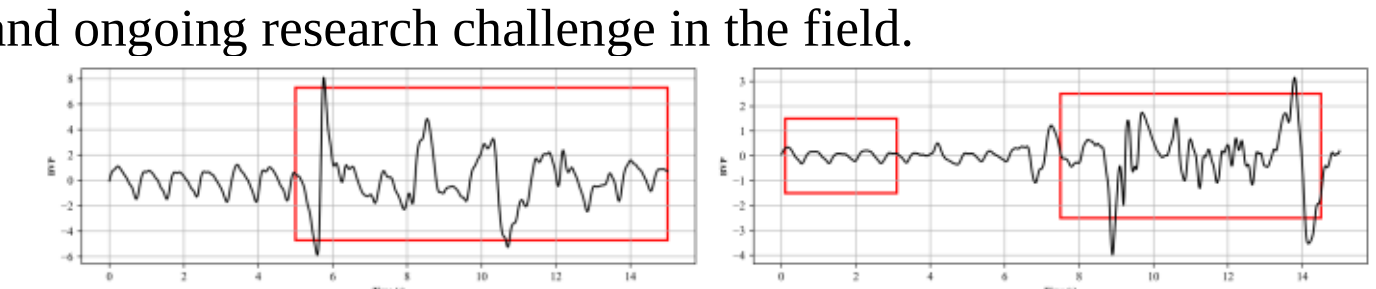

Fig. 4. Illustration of poor-quality signals contributing to reduced performance during LOSO validation. The BVP signal from Subject-5 of the RUET SPML dataset (left) exhibits significant motion artifacts, while the signal from Subject-15 of the WESAD dataset (right) lacks clear systolic and diastolic markers, also suffers from motion artifacts, resulting in degraded signal quality.

### *G. Interpretability of the Proposed Model*

The interpretability of the proposed TEANet model is illustrated through feature visualization using uniform manifold approximation and projection (UMAP) [45] technique, as shown in Fig. 5. The figure provides a two-dimensional representation of the features as they progress through various layers of the model during the learning process. The top and bottom row represent the feature visualization of the RUET SPML (top), and WESAD dataset (bottom) respectively. The red dots represent the features of the normal class, while the blue dots represent the features of the stressed class. The first column represents the feature visualization after the classification block in the TEANet. In this stage, the features of both normal (blue) and stressed (red dots) classes samples are indistinguishable, indicating that the initial stage lack clear separation of the features. The second column illustrates the features after the TEA layer (before the classification block). In this stage, as the features pass through the subsequent blocks (number of layers), the model extracts significant patterns from the signal, leading to a gradual improvement in the separability. Lastly, the third column represents the feature representation of the 1D global average pooling layer of both datasets. The clear distinction in the feature space at this stage underscores the effectiveness of the model in capturing complex patterns and dependencies within the data. This progressive enhancement in feature separability across layers demonstrates the model's capacity to learn meaningful representations, ultimately contributing to its robust classification performance. This visualization conveys the model's ability to refine and distinguish data as it delves deeper into the model, providing a clear and interpretable explanation of the decision-making process within the TEANet.

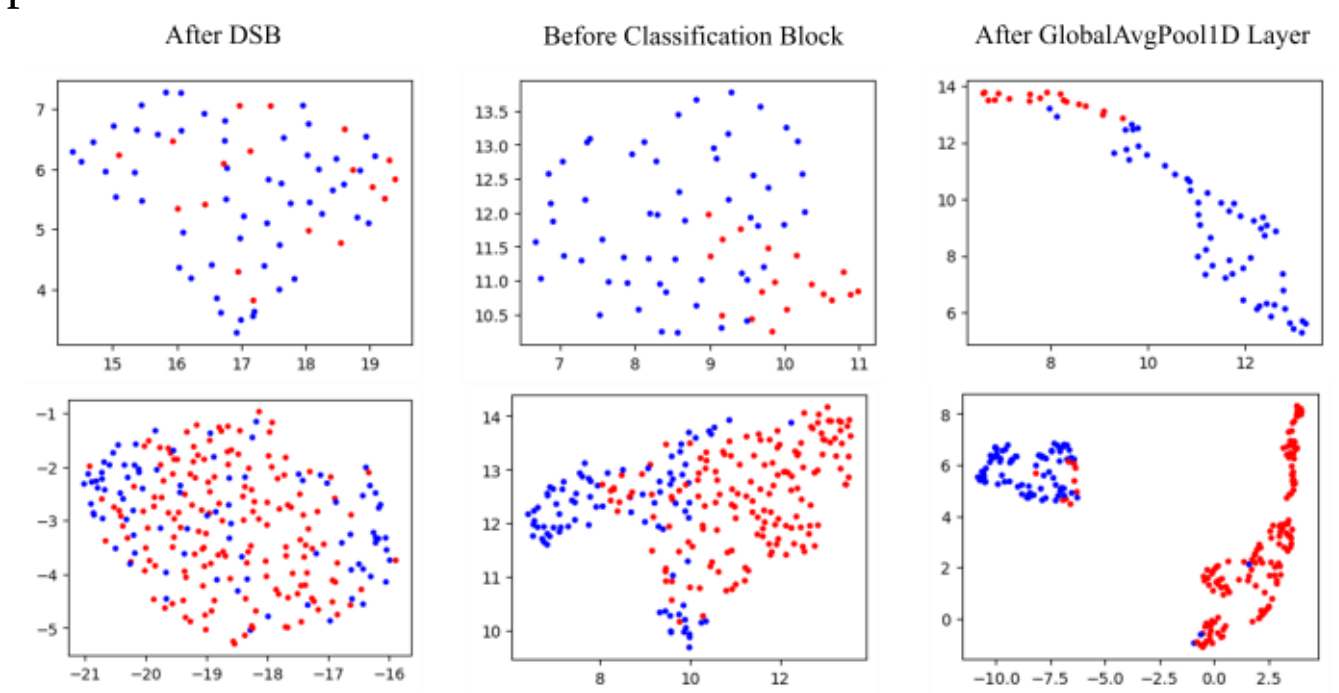

Fig. 5. Feature visualization of various layers illustrating the learning process of the TEANet model using uniform manifold approximation and projection. The top row depicts feature representations from the RUET SPML dataset, while the bottom row shows feature representations from the WESAD dataset. These visualizations highlight the effective class separation between the normal (red) and stressed (blue) classes.

### *H. Explainability of the Proposed Model Using Feature Maps*

Deep learning models are often perceived as black boxes, limiting their applicability in critical domains due to a lack of explainability. Feature maps provide a valuable means to address this challenge by offering insights into the model's decision-making process. Fig. 6 depicts the feature maps of signals processed by the proposed model, emphasizing the regions of the signal that the model identifies as most significant for its decisions. The feature maps were extracted from the last convolutional layer, which contains 32 filters. The teal-colored feature maps (left) represent normal samples, while the purple-colored feature maps (right) correspond to stressed samples of the BVP signals. The distinct activation patterns observed between the normal and stressed classes demonstrate the model’s capacity to differentiate effectively between these states, underscoring its ability to identify critical features for decision-making process.

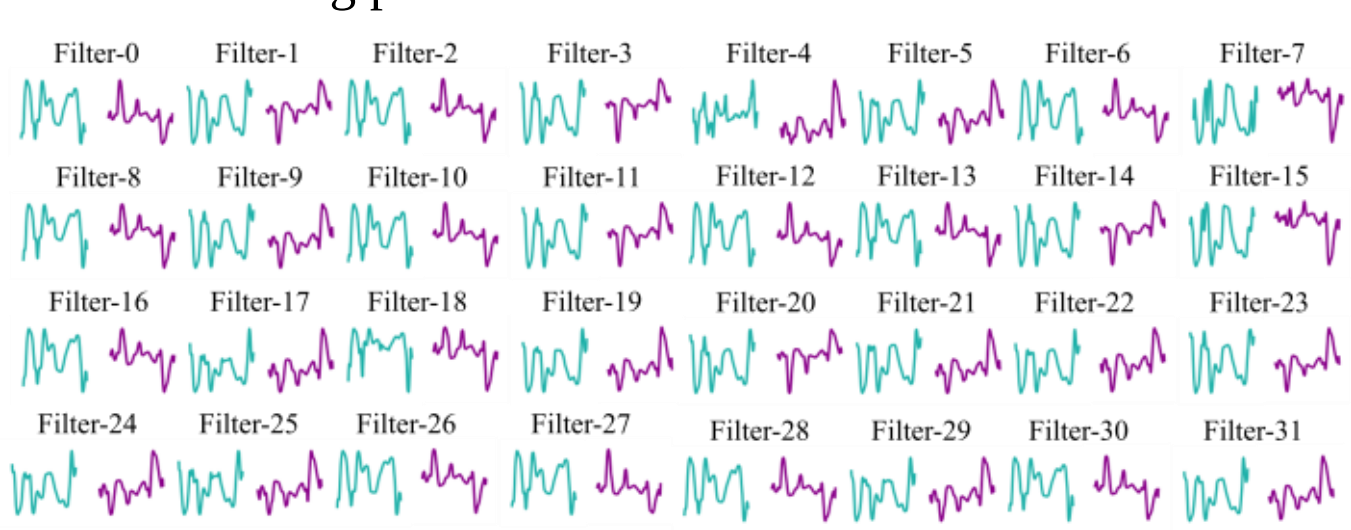

Fig. 6. Feature map of the last convolution layer of the classification block having 32 filters for normal (left) and stressed (right) of each filter on the RUET SPML dataset capturing different patterns of the BVP signal. The teal color represents normal class samples, and the magenta color represents the stressed class samples respectively.

### *I. Explainability of the Proposed Model Using Frequency Band-wise Integrated Gradients Attribution*

To analyze the spectral importance of the signal component, we calculated the integrated gradients (IG) values [46] for three different frequency bands: band 1 (0-2 Hz), band 2 (2-4 Hz), and band 3 (4-10 Hz) referring to the eqn. 14.

$$I_{G_{b_i}} = \int_{\alpha=0}^{1} \frac{\partial f_C(x_{baseline} + \alpha(x(b_i) - x_{baseline}))}{\partial x} d\alpha \qquad (14)$$

As illustrated in Fig. 7, the model's classification of the stress classes primarily relies on the frequency components within band 1, followed by band 2 as the next most influential range. Similarly, for the non-stress classes, band 1 emerged as the most salient frequency band, with bands 2 and 3 also contributing to the prediction. Since the majority of cardiac information is captured within the 0–4 Hz frequency range of the BVP signal, rather than the 4–10 Hz range, these findings suggest that the model effectively emphasizes informative signal components while minimizing the influence of less informative frequency bands in its decision-making process.

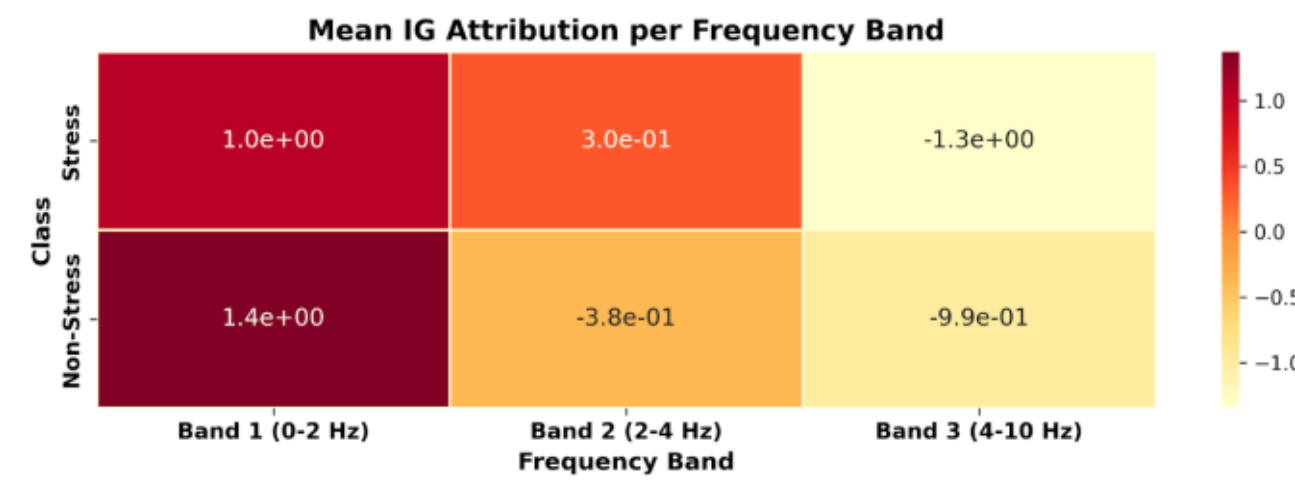

Fig. 7. Frequency band-wise Integrated Gradients attribution for holdout analysis of RUET-SPML dataset, indicating that the frequency band 1 (0–2 Hz) and frequency band 2 (2–4 Hz) have the highest impact on the model’s decision-making process.

### *J. Compatibility with the Resource Constrained Platform*

The model was trained and tested using a Tesla T4 Graphics Processing Unit (GPU). To further assess its suitability for resource-constrained environments, the hold-out validated model was also deployed on a Raspberry Pi 3B+, featuring a 1.4 GHz quad-core Cortex-A53 CPU and 1 GB RAM. The full deployment setup is depicted in Fig. 8.

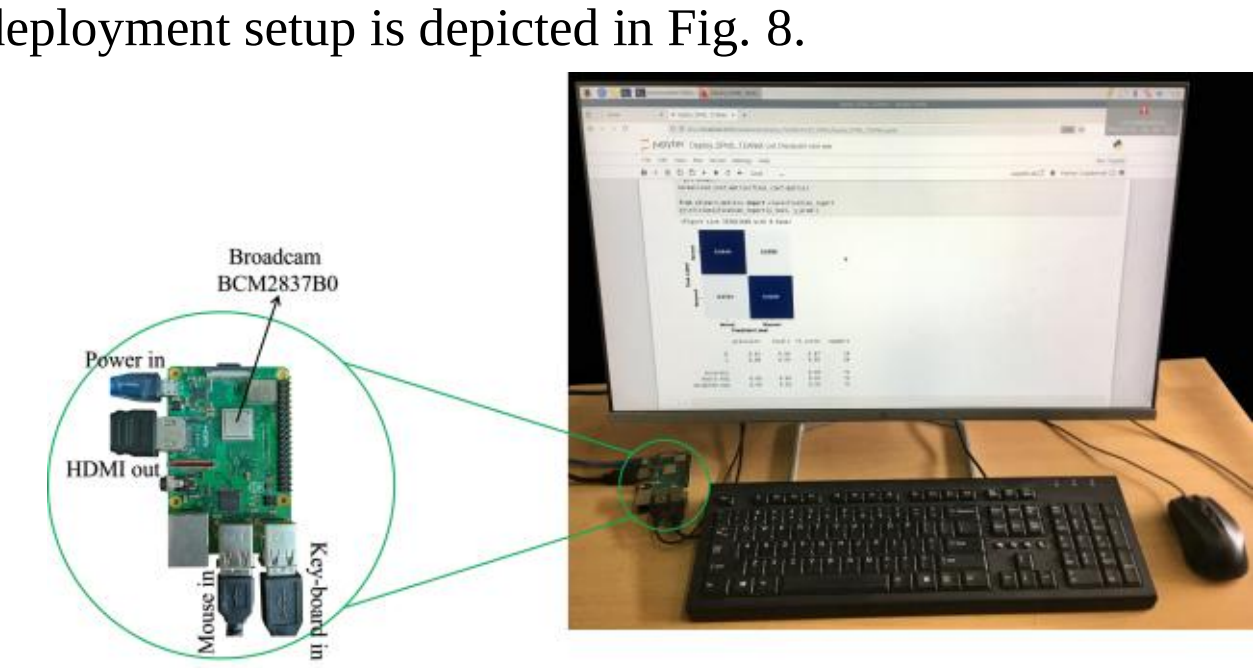

Fig. 8. Deployment of TEANet on a Raspberry Pi 3B+, a resource-constrained device with a 1.4 GHz quad-core CPU and 1 GB RAM.

For deployment of the model on the Raspberry Pi, the model was converted into TensorFlow Lite (TFLite) format while preserving float32 precision to mitigate accuracy degradation. As summarized in Table VII, the TFLite conversion did not lead to any performance drop on the RUET SPML dataset; the accuracy, sensitivity, specificity, F1-score, AUC, and Cohen’s kappa remained identical with and without the TFLite model and augmentation achieving 93.06% accuracy, 94.44% specificity, 92.59% sensitivity, 95.24% F1-score, 0.9835 AUC, and 0.8246 kappa. For the WESAD dataset, a minor variation was observed post-TFLite conversion: the accuracy slightly decreased from 96.94% (without TFLite) to 96.15% (with TFLite), and the F1-score dropped from 95.59% to 94.52%. This marginal reduction is likely due to runtime overhead and hardware constraints on the Raspberry Pi. Nonetheless, the TFLite model maintained robust performance, achieving over 93% accuracy on RUET SPML and over 96% on WESAD demonstrating the feasibility of deploying the TEANet model on low-power wearable devices for real-time stress detection.

TABLE VII

PERFORMANCES OF THE TEANET WITH AND WITHOUT CONVERSION INTO TENSORFLOW LIGHT.

| Dataset | TFlite | Acc | Spe | Sen | F1 | AUC | k |
|---|---|---|---|---|---|---|---|
| RUET SPML | Yes | 93.06 | 94.44 | 92.59 | 95.24 | 0.9835 | 0.8246 |
| | No | 93.06 | 94.44 | 92.59 | 95.24 | 0.9835 | 0.8246 |
| WESAD | Yes | 96.15 | 97.20 | 94.20 | 94.52 | 0.9798 | 0.9156 |
| | No | 96.94 | 97.06 | 96.63 | 95.59 | 0.9824 | 0.9350 |

## IV. Conclusion

This work proposed a novel model, TEANet, for automated stress monitoring using wearable BVP signals. Given the non-invasive nature and widespread availability of BVP sensors, the proposed model effectively addresses the critical need for efficient stress assessment in resource-constrained environments. The model's performance was comprehensively evaluated across diverse datasets, demonstrating superior results compared to existing state-of-the-art methods. This paper highlights the model's robustness, achieving the highest accuracy of 92.94% and an F1 score of 95.16% using the RUET SPML dataset, and a highest accuracy of 96.94% and an F1 score of 95.95% for stress detection using the WESAD dataset, evaluated through leave-one-subject-out cross-validation. This validation approach underscores the model's generalizability for real-world applications. A key contribution of this work is its demonstration of superior stress monitoring performance using only a single BVP signal, which is typically considered less informative for stress analysis. This underscores the robustness and practicality of the model, particularly given the ease of acquiring BVP signals in wearable devices. Moreover, UMAP-based feature visualization across different layers is utilized, offering insights into the model's decision-making process and improving interpretability. Additionally, integrated gradient attribution based explainability is applied to further explain the model’s transparency in decision-making across different frequency bands. To evaluate the model’s practical viability, it was also deployed on a resource-constrained device for real-time stress classification. The results demonstrate the model’s strong potential for integration into wearable devices, enabling continuous and non-invasive stress monitoring in real-world settings. Future work will focus on extending the model to multi-level stress classification and incorporating additional physiological signals using short-duration segments.